\documentclass[letterpaper]{jpconf}
\usepackage{graphicx}
\usepackage{amsmath}
\begin{document}

\title{Non-Markovian Dynamics of Charge Carriers in Quantum Dots}

\author{Eduardo Vaz and Jordan Kyriakidis}

\address{Department of Physics, Dalhousie University, Halifax, Nova
  Scotia, Canada, B3H~3J5}

\ead{jordan.kyriakidis@dal.ca}

\begin{abstract}
  We have investigated the dynamics of bound particles in multilevel
  current-carrying quantum dots.  We look specifically in the regime
  of resonant tunnelling transport, where several channels are
  available for transport.  Through a non-Markovian formalism under
  the Born approximation, we investigate the real-time evolution of
  the confined particles including transport-induced decoherence and
  relaxation.  In the case of a coherent superposition between states
  with different particle number, we find that a Fock-space coherence
  may be preserved even in the presence of tunneling into and out of
  the dot.  Real-time results are presented for various asymmetries of
  tunneling rates into different orbitals.
\end{abstract}

\section{Introduction}

Some of the most peculiar features of quantum theory, such as the
existence of quantum superpositions and of entangled states, are
typically destroyed by uncontrolled and ultimately inevitable
interactions with a surrounding, often incoherent
environment~\cite{breuer_OQS}.

Most of the recent theoretical analysis done in the area of
low-dimensional dynamical quantum systems has been either for open
Markovian systems, where the past memory of the system is neglected,
or for closed unitary systems, where the dynamics are reversible.
Furthermore, much of the research done until recently has focused on
steady state phenomena where a Markovian approach can be expected to
provided reasonable results.  

However, the \emph{transient} behavior of the system carries a
tremendous amount of information in the form of coherence and
relaxation dynamics.  Ultrafast laser pulse excitations, for example,
are providing insight into the heterostructure dynamics on a
femtosecond timescale~\cite{shah_1999}.  A theory which accounts for
quantum dynamical behavior on the same time scale is therefore of
great benefit not only to experiment, but also to basic understanding.

The formalism of non-equilibrium Green's functions (NEGF) and
formalisms based on NEGF, have been extraordinarily successful in the
analysis of much phenomena in mesoscopic 
systems~\cite{kadanoff_baym,haug_koch,haug_jauho}.   
Notably, work in both elastic and inelastic transport in quantum dots 
has been successfully done by means of NEGF or derivatives of this 
formalism \cite{wacker_jauno_1998,wacker_jauno_1999}.  However, 
NEGF is inherently a closed-system formalism
where the system has Hamiltonian dynamics~\cite{datta_transport}, and
thus does not account for irreversibility arising from interactions
with an unseen, unknown, or otherwise intractable environment.
Promising attempts have been made to extend NEGF to open quantum
system, for example, by treating the environment as a correction to
the system's self-energy~\cite{datta_2000}, or by calculating two-time
correlation functions~\cite{Knezevic:2003p3439}, effectively
separating the time scales into transient and steady-state regimes.
These and other approaches constitute significant progress toward a
full non-Markovian open system analysis within the NEGF formalism.

Even so, if the evolution of the (possibly correlated) many-body
system states themselves are sought, one must calculate the Green's
functions and then extract the density matrix, which in the
non-equilibrium case may not be unique \cite{Knezevic:2003p3439}.  
An alternative formulation, one we adopt in the sequel, is
to work with the density matrix directly.  This arguably provides a
more intuitive description of the dynamics of non-Markovian open
quantum systems, and it readily yields the actual states of the system
as well as the quantum coherence between them.

The focus of this paper is to thus present a generic formalism to
investigate the dynamics of a few-electron quantum dot in a fully
non-Markovian fashion in the density matrix formalism.
Section~\ref{sec:evol-dens-matr} presents a brief outline of the
derivation of the evolution equations for the density matrix of the
system in both the Markovian and non-Markovian regimes.
Section~\ref{sec:model} presents the model Hamiltonian for our system
as well as the analytical expression for the transition tensor for
this model.  Section~\ref{sec:two-channel-four} illustrates the
non-Markovian approach by considering the case of a multilevel dot in
a regime where two channels are available for transport.  Results are
presented for the transient population probabilities in regimes where
the tunnelling strength between system and leads varies from orbital
to orbital.  In Sec.~\ref{sec:fock-space-coherence} we give a brief
discussion on Fock-space coherences and their implications for Quantum
Information.  Finally, in Sec.~\ref{sec:conclusion}, we present a
summary and conclusion.

\section{Evolution of the density matrix for an open quantum system}
\label{sec:evol-dens-matr}

In isolated quantum systems, the time evolution can be described by a
single state vector $|\psi(t)\rangle$.  Once the system is coupled to
a larger environment whose degrees of freedom are either unknown,
intractable, or uninteresting, the environmental degrees of freedom
may be traced out and the resulting state of the system alone is
described not by a state vector, but by a density operator~\cite{blum}
$\rho(t) = \sum_n W_n(t) | \psi_n(t)\rangle \langle \psi_n(t) |$,
where $W_n(t)$ is the probability that state $|\psi_n(t)\rangle$ is
occupied at time $t$.  The time-dependence of the density operator can
be obtained by an iterative integration of the Liouville-von~Neumann
equation, which gives
\begin{equation}\label{eq:liouville_total}
  \dot \rho(t) =  -\mathcal{L}(t)\rho(0) + \int_0^t \! dt'
  \mathcal{L}(t) \mathcal{L}(t') \rho(t') +
  \int_0^t \! dt' \! \int_0^{t'} \!dt'' 
  \mathcal{L}(t)\mathcal{L}(t')\mathcal{L}(t'') \rho(t'') + \ldots,
\end{equation}
where $\mathcal{L}(t)$ is a Liouville superoperator such that,
$\mathcal{L}(t)F = i \hbar^{-1}[H(t),F]$ for any operator $F$ and
Hamiltonian $H(t)$.

Equation~\eqref{eq:liouville_total} represents the evolution of the
complete system plus environment.  Under the Born approximation, the
evolution equations are truncated at second order in the interaction
Hamiltonian.  After some manipulation~\cite{blum}, the resulting
equations for the matrix elements of the density operator read,
\begin{equation}
  \label{eq:rho_dot_redfield}
  \dot \rho_{ab}(t) = \sum_{c,d} \int_0^t \! dt'
    \rho_{cd}(t') R_{abcd}(t - t') e^{i\gamma_{abcd} t'},
\end{equation}
where, $R_{abcd}(t - t')$ is a Redfield-type transition tensor
containing all the information characterizing the system, the
environment, and the coupling between them, and where $ \gamma_{abcd}
= \omega_{ab} - \omega_{cd} = (E_a - E_b + E_d - E_c) / \hbar$ denotes
energy differences between system states.

For an environment with a much greater number of degrees of freedom
relative to those of the system, and for a weak enough
system-environment coupling, we make the assumption that the system
has no effect (no back-action) on the environment, that the system is
not correlated with the environment, and that the environment is and
remains in equilibrium~\cite{blum}.  In this case, the total density
matrix for the system plus environment can be written as,
\begin{equation}
  \rho(t) \approx \rho_{\text{system}}(t) \varrho_{\text{R}}(0). 
\end{equation}
where $\rho_{\text{system}}(t)$ is the density matrix for the open
quantum system, and $\varrho_{\text{R}}$ is the density matrix for the
environment.

\subsection{Markovian Approach}
\label{sec:markovian-approach}

The Markov approximation assumes an environment which relaxes on a
time scale time much faster than the characteristic time of the
system.  The environment correlation functions vanish at such a fast
rate, that the reversibility of the system is essentially destroyed,
thus also destroying the memory of the system~\cite{breuer_OQS}.
Thus, $\dot \rho_{\text{system}}(t)$ depends only on the present value
of $\rho(t)$.  That is, it becomes local in time.  For time intervals
$t-t'$ much greater than the environment's correlation time $\tau$,
the correlation functions for environment operators rapidly become
uncorrelated and decay to zero, $ \langle F_k^\dag (t-t')F_{k'}\rangle
\approx \langle F_k^\dag(t-t')\rangle \langle F_{k'}\rangle \approx
0$.  In the limit $t-t'>>\tau$ the upper integration limit
in~\eqref{eq:rho_dot_redfield} can be extended to infinity with
negligible error in the calculations.  Since the time when the
coupling is turned on is arbitrary, the lower integration limit can be
taken to negative infinity, and the coupled set of
integro-differential evolution equations become a coupled set of first
order ordinary differential equations:
\begin{equation}
  \label{eq:rho_markov}
  \dot \rho_{ab}(t) \longrightarrow \sum_{cd} \rho_{cd}(t)
  \int_{-\infty}^{\infty} \! dt' R_{abcd}(t - t') 
  e^{i\gamma_{abcd} t'} = \sum_{cd} \rho_{cd}(t) W_{abcd}.
\end{equation}
This in turn leads to a Fermi's Golden rule for transitions with
strict energy conservation for all transitions.

Although much of the recent theory in the area of non-equilibrium
quantum dot systems has made use of Markovian-type
approximations~\cite{Apalkov:2007p541,Imura:2007p227,pedersen:235314,Egorova:2003p212,Legel:2007p896,Vaz:2006p587},
the approach has fairly severe consequences on the transient behavior
of the system, especially when considering coherence properties of the
system at short time scales.  Nonetheless, the Markov approximation
may be expected to yield reasonable results at sufficiently long
times~\cite{blum}, when the system has reached a steady state, or when
considering intermediate time scales of averaged behavior, such as the
average total current through a system.

\subsection{Non-Markovian Approach}
\label{sec:non-mark-appr}

In order to account for the transient dynamics, especially for the
characteristics of the coherence between system states, memory effects
must be preserved.  In our approach we make use of the convolution
theorem for Laplace transforms, to represent
Eq.~\eqref{eq:rho_dot_redfield} as a set of algebraic equations in
Laplace space,
\begin{equation}
  \label{eq:rho_dot_redfield_laplace}
  s \rho_{ab}(s + i \omega_{ab}) -\rho(0) = \sum_{cd} 
  \rho_{cd}(s + i \omega_{cd}) R_{abcd} (s + i \omega_{ab}).
\end{equation}
This set of coupled algebraic equations can be solved analytically for
$\rho_{ab}(s)$ for only a few system states ($\sim 2$--4), since the
computational effort rapidly increases with the number of available
transport channels.  In the regime of sequential resonant tunnelling
transport through a quantum dot containing non-interacting electrons,
the transport channels are those single-particle system states
$|\alpha \rangle$, whose energy lies within the bias window.  These
single-particle states define the possible dynamical many-body states
of the system as those involved when the single particle states are
empty or occupied.  Thus, for $k$ channels, a minimum of $2^k$
many-body system states are required (for empty or occupied).
Furthermore, $(2^k)^2$ density-matrix elements are required do
describe the population probabilities as well as the coherence between
the states.  The temporal evolution of the density-matrix elements
themselves  are governed by $(2^k)^4$ transition tensor elements
(Redfield-type in the Born approximation).  For larger sizes, a
numerical approach may be utilised to invert the linear system in
Eq.~\eqref{eq:rho_dot_redfield_laplace} at each (complex-valued) point
$s$, thus obtaining $\rho_{ab}(s)$.  The solutions $\rho_{ab}(s)$ in
Laplace space can then be brought back to time space by applying an
inverse Laplace transform.  This is done numerically by performing a
Bromwich integral~\cite{abramowitz+stegun} over a suitably chosen
contour.  For every point $s$ along the Bromwich contour, a matrix
inversion is performed to evaluate all components of $\rho_{ab}(s)$.
Such a Bromwich integration is (numerically)
performed~\cite{integration} at each time step, thus obtaining
$\rho_{ab}(t)$.

\subsection{Violation of Positivity}
\label{sec:violation-positivity}

It has been established by Kraus~\cite{Kraus:1971p2939},
Lindblad~\cite{Lindblad:1976p2837}, and Gorini, Kossakowki, and
Sudarshan~\cite{Gorini:1976p2990}, that the evolution of an open
quantum system must remain completely positive to ensure the physical
validity of the states of the system at all times.  However, this
positivity may be broken by introduction of ad hoc relaxation rates,
by the choice of basis states, or by the truncation of higher-order
interaction terms in the evolution equations (such as in the Born
approximation), which can yield serious inconsistencies such as
negative probabilities.  Despite this, much of the investigations of
non-Markovian dynamics acknowledge that for a set of initial
conditions, the theory will reproduce physically consistent results.
Our findings agree with this premise~\cite{vaz_nonMarkov}.

\section{Model}
\label{sec:model}

%\subsection{Hamiltonian}
We model our system as a single generic quantum dot coupled to a
source and a drain reservoir by a tunneling Hamiltonian~\cite{Mahan}.
The total Hamiltonian is given by,
\begin{subequations}
  \label{eq:model}
  \begin{equation}
    \label{eq:mainHamil}
    H = H_S + H_{QD} + H_D + H_T,
  \end{equation}
  where $H_S$ and $H_D$ are the source and drain Hamiltonians,
  respectively.  These are taken to be non-interacting Fermion systems
  shifted by the bias:
  \begin{equation}
    \label{eq:HReservoir}
    H_{S(D)} = \sum_{s(d)} (\epsilon_{s(d)} \pm \frac{1}{2} eV_B) 
    d^\dag_{s(d)} d_{s(d)},
  \end{equation}
  with $d^\dag_{s(d)}$ a creation operator for the source (drain), and
  $d_{s(d)}$ an annihilation operator.

  The Hamiltonian for the quantum dot in Eq.\eqref{eq:mainHamil} is given by,
  \begin{equation}
    \label{eq:QDotHamil}
    H_{QD} = \sum_i (\hbar\omega_i + eV_g) c^\dag_i c_i + V_{\mathrm{int}},
  \end{equation}
  where the single-particle energies $\hbar \omega_i$ are all shifted by
  the applied gate voltage, and $V_{\mathrm{int}}$ is the interaction
  among the confined particles.

  The tunneling Hamiltonian describing the reservoir-dot coupling is given by,
  \begin{equation}
    \label{eq:TunnelHamil}
    H_T = \sum_{k, r = (s, d)} \left( T^r_k d^\dag_r c_k + h.c \right),
  \end{equation}
\end{subequations}
where $T^{s(d)}_k$ is an energy-independent tunneling coefficient for
a particle tunneling from the single-particle state $|k\rangle$ in the
dot to the source (drain) reservoir.  
Finally, we assume zero temperature such that the reservoirs are in
their respective non-degenerate ground states.

%\subsection{The Transition Tensor}

As mentioned above, all system and environment information, including
the characteristics of the interaction between them are contained in
the memory kernel, $R_{abcd}(\tau)$ appearing in
Eq.\eqref{eq:rho_dot_redfield}.  For the model in
Eq.~\eqref{eq:model}, we obtain
\begin{multline}
  \label{eq:redfield_tensor}
  R_{abcd}(\tau) = \sum_{\alpha,\beta,R}
  K^R_{\alpha\beta}(\tau) \left \{
    \Omega^{\alpha}_{\phi_B,\mu^R}
    \Delta^{\alpha\beta}_{badc}-
    \Omega^{\alpha}_{\phi_T,\mu^R}
    \Delta^{\alpha\beta}_{cdab} \right \} \\
  + K^R_{\beta\alpha}(\tau) \left \{
    (\Omega^{\alpha}_{\phi_B,\mu^R} )^*
    \Delta^{\alpha\beta}_{abcd}-
    (\Omega^{\alpha}_{\phi_T,\mu^R})^*
    \Delta^{\alpha\beta}_{dcba} \right \}, 
\end{multline}
where the indices $a,\ b,\ c,\ \text{and } d$ denote many-body states,
$\alpha, \text{ and } \beta$ are denote single-particle states, and $R
= S,\ D$ runs over the source and drain leads.  The energy
$\phi_{T(B)}$ denotes the top (bottom) of the band, and $\mu^R$ is the
chemical potential of reservoir $R$.  Furthermore, we have defined
\begin{subequations}
  \label{eq:redfield_parameters}
  \begin{gather}
    \Omega^{\alpha}_{x,y} \equiv e^{i
      \omega_{x,\alpha}\tau}-e^{i\omega_{y,\alpha}\tau},\\
    \Delta^{\alpha,\beta}_{abcd} \equiv \langle a|
    c^{\dag}_{\alpha}|c\rangle \langle d| c_{\beta}|b\rangle - \langle
    a|c \rangle \langle b| c_{\alpha} c^{\dag}_{\beta}|d\rangle,
  \intertext{and}
    K^R_{\alpha,\beta}(\tau) \equiv \frac{i N_R}{\hbar \tau}\left (
      T^*_{\alpha} T_{\beta} \right )_R,
  \end{gather}
\end{subequations}
where $\omega_{x,\alpha}=(E_x - E_\alpha) / \hbar$ is a frequency
denoting the energy difference between states $|x\rangle$ and $|\alpha
\rangle$, $T_{\alpha}$ is the tunneling coefficient to a single
particle state $|\alpha \rangle$ in the dot, $N_R$ is the density of
states for the 2DEG environment, and $\tau$ is a time parameter.  

With this general form for the transition tensor, given a set of basis
states, the individual elements of the tensor are calculated.  The
form of the Redfield transition tensor allows for independent tuning
and analysis of barrier width symmetries, and of asymmetries between
core-states and edge-states couplings to the reservoirs, by means of
the independent tunneling rates.

\section{Two-channel (four-state) system}
\label{sec:two-channel-four}

\begin{figure}[b]
\begin{center}
  \includegraphics[width=24pc]{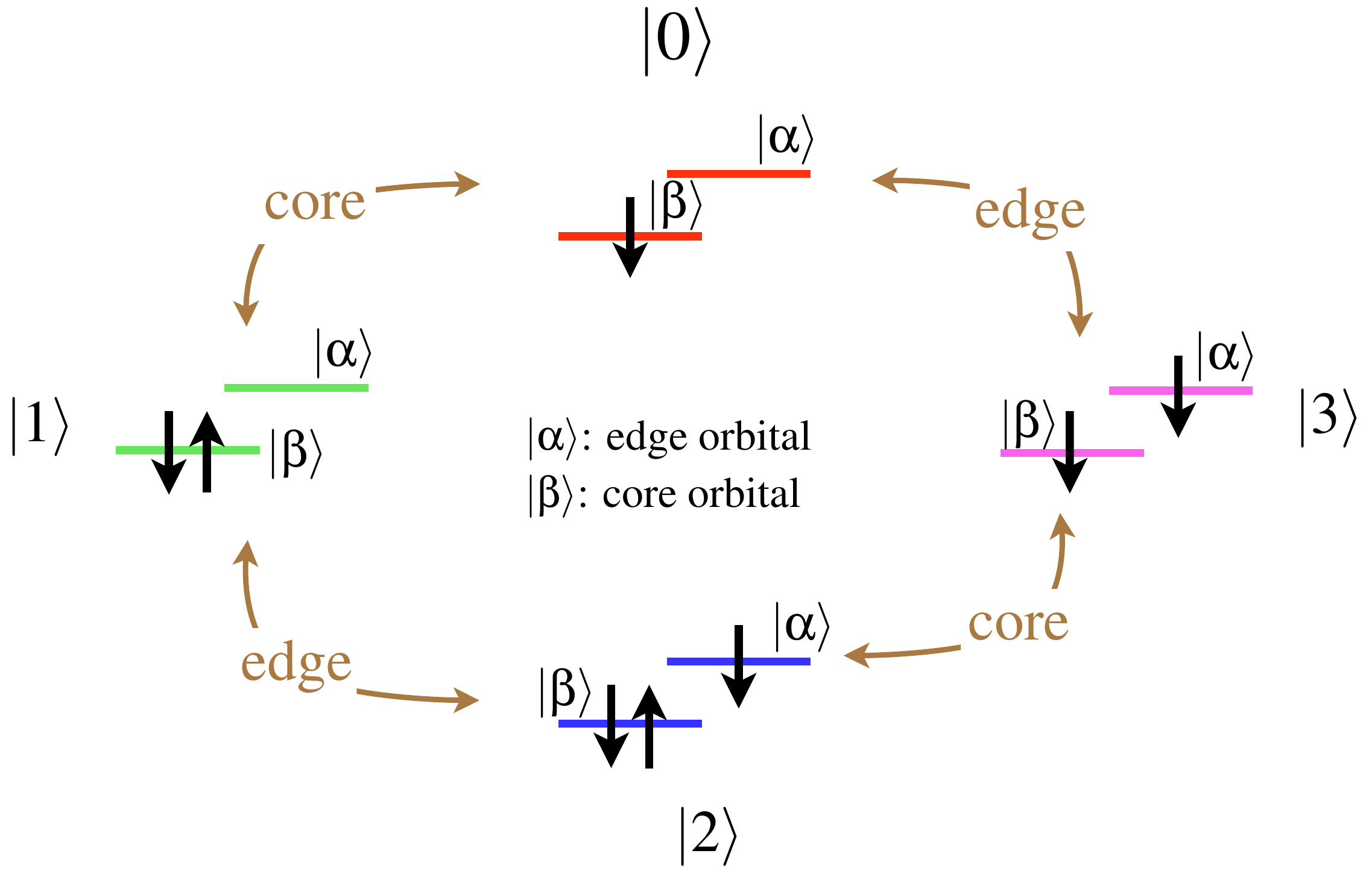}\hspace{2pc}%
  \begin{minipage}[b]{11pc}
    \caption{\label{fig:schem-repr-syst}Schematic representations of
      system under consideration.  Four system states are considered
      depending on the relative occupation of the single-particle
      orbitals $|\alpha\rangle$ and $|\beta\rangle$, each of which can
      in principle have different tunnel couplings to the leads.}
  \end{minipage}
\end{center}
\end{figure}

As an illustrative example, we present results for the diagonal
elements of $\rho(t)$ in the above model.  For definiteness, we
consider a quantum dot with $N$ confined particles, and with two
resonant tunnelling transport channels available within the bias
window.  Each channel involves a particle-number fluctuation between
$N$ and $N \pm 1$, and can involve either the ground or first excited
state of the $N$-particle system.  In general, the availability of
$k$ transport channels involves a minimum of $2^k$ states.  In the
present case, the four relevant states are
\begin{equation}
  \label{eq:states}
  |0\rangle = |(N-1)_{g.s.} \rangle,\quad
  |1\rangle = |N_{g.s.}\rangle,\quad
  |2\rangle = |(N+1)_{g.s.}\rangle,\quad
  |3\rangle = |(N)_{e.s.}\rangle.
\end{equation}
Here, the state $|0\rangle$ denotes the $(N-1)$-particle ground state
of the system, $|1\rangle$ and $|2\rangle$ denote ground-states of the
$N$ and $(N+1)$-particle system respectively, and $|3\rangle$ denotes
the first excited state of the $N$-particle system.  A schematic
representation of the system is shown in
Fig.~\ref{fig:schem-repr-syst}.

Each orbital will in general be coupled differently to the leads,
owing to the detailed shape of the
wave function~\cite{ciorg02:collap.spin.singl}.  We identify in
Fig.~\ref{fig:schem-repr-syst} edge and core orbitals.  Core orbitals
are weakly coupled to the leads owing to a poor ($s$-type) overlap
with the lead states whereas edge orbitals ($p$-type) are more
strongly coupled to the leads.

\begin{figure}[t]
\begin{center}
  \includegraphics[width=6in]{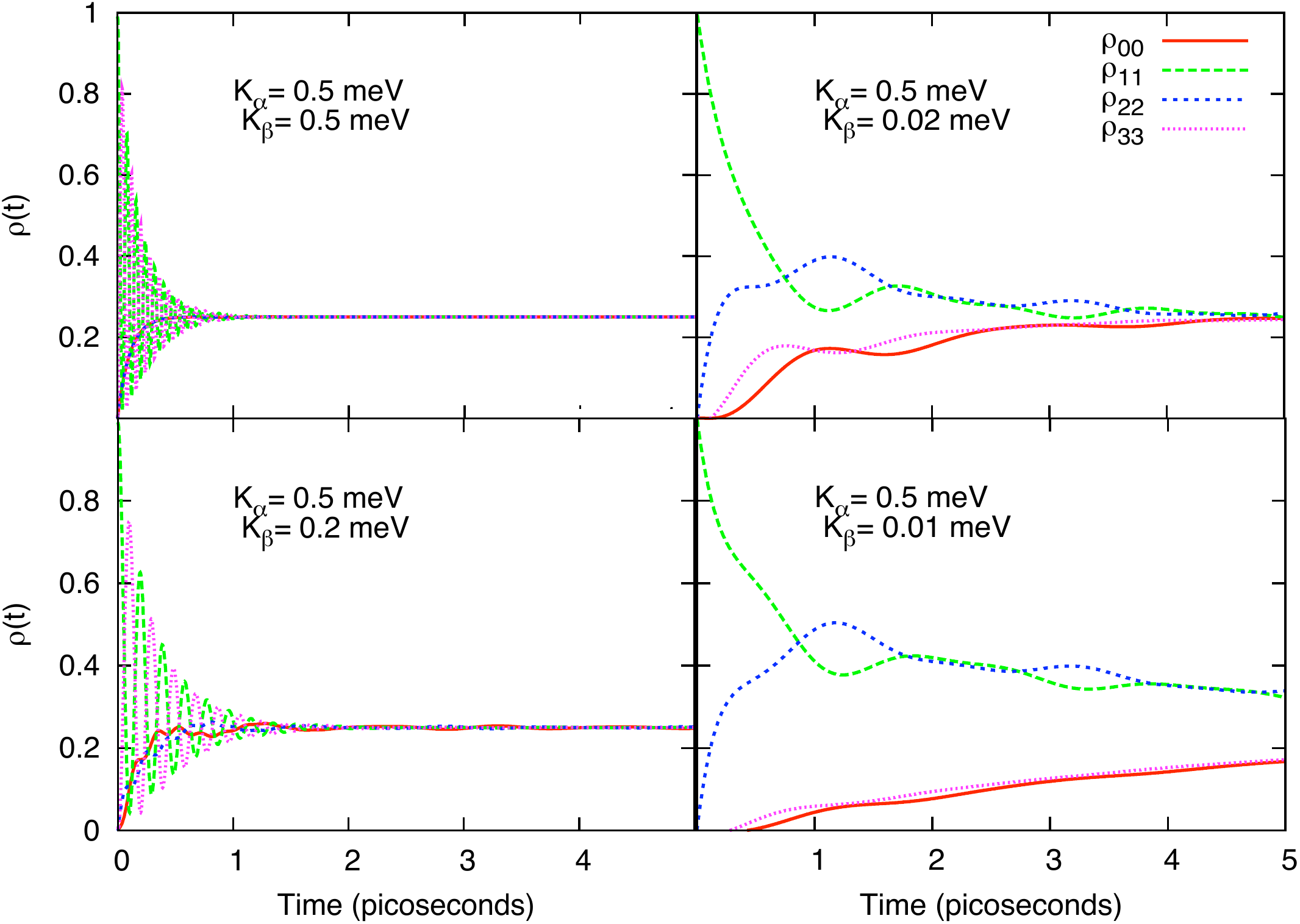}
  \caption{\label{fig:nonMarkov}Time evolution of population
    probabilities in a quantum dot with 2 transport channels and four
    states.  (See Eq.~(\ref{eq:states}).)  The plots are for symmetric
    source and drain tunnel barriers, and varying core edge asymmetry.
    We assume a 6~meV bias symmetric about the Fermi energy, and two
    transport channels at energies $epsilon_{\text{Fermi}} \pm
    1$~meV.}
\end{center}
\end{figure}

Figure~\ref{fig:nonMarkov} shows the real-time evolution of the
population probabilities of the system $\rho_{nn}(t)$ in the large
bandwidth limit.  Results are shown for a bias of $V_{\text{bias}} =
6$~meV, symmetric about the Fermi energy $\epsilon_{\text{Fermi}}$.
We assume two channels within the bias window, $E_{\alpha} =
\epsilon_{\text{Fermi}} + 1$~meV and $E_{\beta} =
\epsilon_{\text{Fermi}} - 1$~meV.
Finally, we keep the barriers fully symmetric (i.e. $T^{\text{source}}
= T^{\text{drain}}$), while the ratio between tunneling to core states
and tunneling to edge states, is varied (i.e. $T^{\alpha}/T^{\beta} >
1$).  Several features are evident in Fig.~\ref{fig:nonMarkov}, and we
discuss three in particular.

First, we see that for the smallest couplings to the core orbitals
(the two plots on the right of Fig.~\ref{fig:nonMarkov}), the four
levels couple into two distinct pairs.  States $|1\rangle$ and
$|2\rangle$---the $N$ and $(N+1)$-particle ground states---are
strongly coupled, as are states $|0\rangle$ and $|3\rangle$---the
$(N-1)$-particle ground state and the $N$-particle first-excited
state.  We can understand why this occurs with recourse to
Fig.~\ref{fig:schem-repr-syst}.  The transition between states
$|1\rangle$ and $|2\rangle$ is through an edge orbital and this
coupling is stronger than the transition between $|1\rangle$ and
$|0\rangle$ which involves a core orbital.  Similar arguments apply
for the coupling between states $|3\rangle$ and $|0\rangle$ (strong
coupling) and between $|3\rangle$ and $|2\rangle$ (weak coupling).

Second, in the steady state ($t \rightarrow \infty$), all occupation
probabilities tend to the same value of 1/4.  This is seen regardless
of the tunnelling strengths of core and edge states.  The equal
probability of 1/4 for each level can be understood as being due to
the symmetric barriers between the dot and the source on the one hand,
and between the dot and the drain on the other.  Since these barriers
are symmetric, any level will have an equal probability of being
either occupied or unoccupied.

Third, although all four states are equally occupied in the steady
state, the time to \emph{actually reach} the steady states does depend
on the relative couplings to the core and edge orbitals.  The fact
that the time taken to reach the steady state increases as the
tunnelling to the core state decreases can be understood as the effect
of decreasing the available tunnelling pathways.  Fewer pathways
available means that the system takes longer to reach the steady
state.  In the limit of zero tunnelling to the core state, for
example, states $|3\rangle$ and $|0\rangle$ will \emph{never} become
occupied.  In this case, the remaining two levels will each reach an
occupation probability of 1/2 rather than 1/4.

\section{Fock-Space Coherence}
\label{sec:fock-space-coherence}

Finally, we comment on the various forms of coherence available to
systems like the present one in which the particle number fluctuates.
By Hilbert-space coherence, we denote coherence between states with
identical particle numbers---states $|1\rangle$ and $|3\rangle$ in
Eq.~(\ref{eq:states}), for example.  In contrast, a Fock-space
coherence can exist between states of differing particle number.  In
the system considered above, we can have a Fock-space coherence
between states differing by one particle---four possibilities for the
states in Eq.~(\ref{eq:states})---or by two particles---states
$|0\rangle$ and $|2\rangle$.  In general, the interaction and
interplay between these distinct classes of coherence is governed by
the Redfield tensor $R_{abcd}(t)$.

\begin{figure}[tb]
\begin{center}
  \includegraphics[width=5.0in]{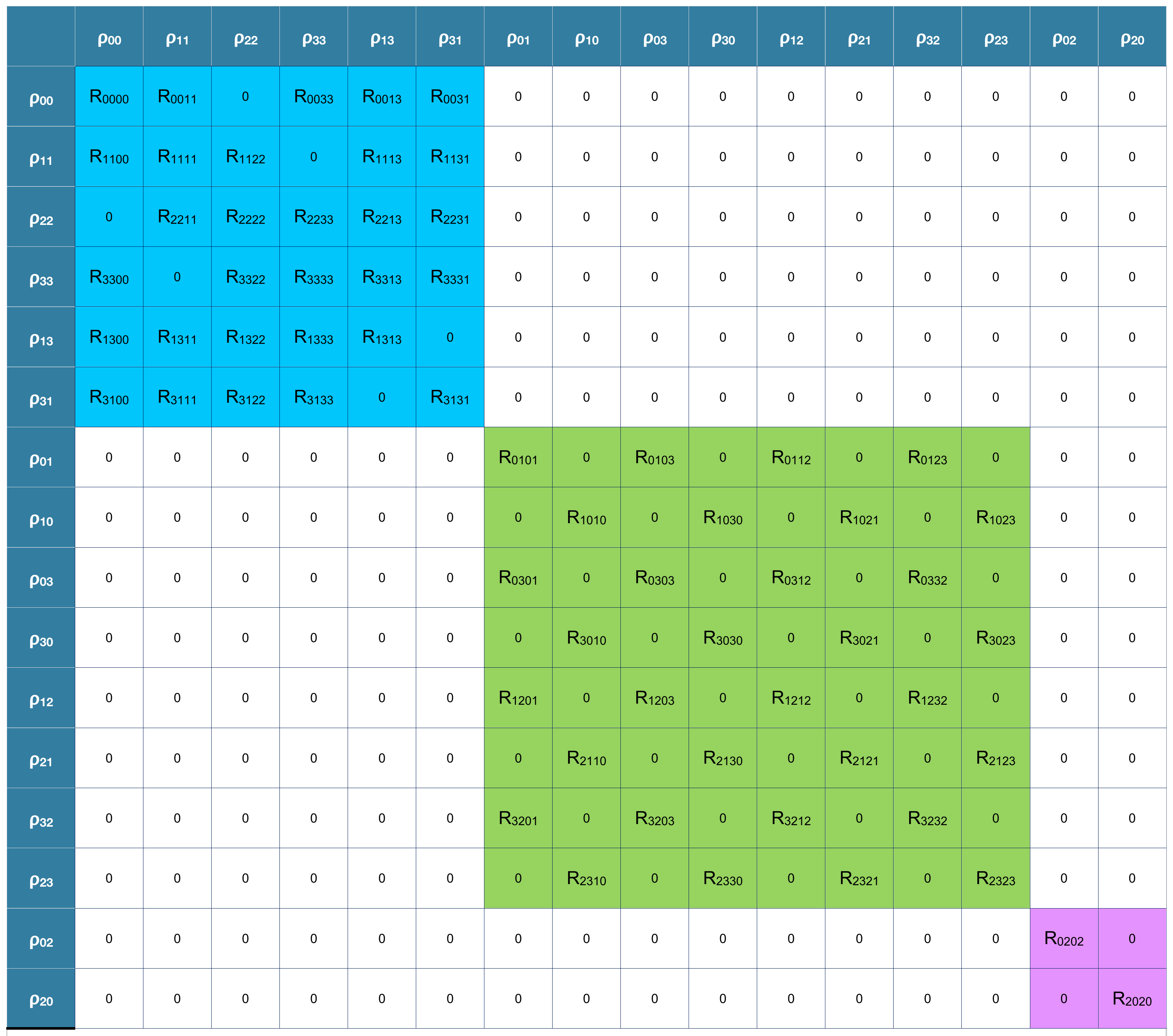}
  \caption{Representation of the non-zero Redfield tensor elements for
    the non-interacting, non-Markovian theory for the 2
    transport-channel case with the four-states: $|0\rangle =
    |(N-1)_{g.s.}\rangle$, $|1\rangle = |N_{g.s.}\rangle$, $|2\rangle
    = |(N+1)_{g.s.}\rangle$, and $|3\rangle = |(N)_{e.s.}\rangle$.
    Three distinct decoupled submatrices become apparent: The top
    submatrix represents the transitions between the population
    probabilities and the coupled coherence between them
    (Hilbert-space coherence), the middle submatrix represents
    transitions relating states with particle numbers differing by 1,
    and the bottom submatrix (a 1 by 1 matrix) represents a transition
    relating states with particle numbers differing by 2.}
\label{fig:plot4}
\end{center}
\end{figure}

For the $k = 2$ channel system, $k^2 = 4$ four states are required, as
outlined above.  This yields $(k^2)^2 = 16$ density matrix elements,
and thus, $(k^2)^4 = 256$ Redfield tensor components, many of which
are not independent and many of which vanish.  In Fig.~\ref{fig:plot4}
we show a graphical representation of all non-zero components of the
Redfield tensor, for the same 2-channel case described above for the
model Eq.~(\ref{eq:model}).  In this matrix representation, we see
three distinct and decoupled submatrices.  The top submatrix
represents the transitions between the population probabilities and
all Hilbert-space coherence terms.  The remaining two block describe
the Fock-space coherence between states differing by either one
(middle submatrix) or two (bottom submatrix) particles.

The remarkable structure evident in Fig.~\ref{fig:plot4} implies that
each block undergoes its own independent time evolution.  In turn,
this suggests that, should a Fock-space coherence be established in
the system at any time, this coherence can be robust and long-lived
even in the rather severe perturbation of transport carriers being
continually injected and removed from the system.  This may have very
promising consequences for quantum information, where strong,
accessible, and long-lived coherence is an essential prerequisite for
information processing beyond the classical limit.

\section{Conclusion}
\label{sec:conclusion}

We have developed a generic non-Markovian formalism for the real-time
evolution of the density matrix of a quantum dot weakly coupled to
source and drain reservoirs.  Within the tunneling Hamiltonian
approach, we have analytically derived an explicit transition
(Redfield) tensor describing the full dynamics of the system.  In the
case of two transport channels (four states) we observed the real-time
evolution of the population probabilities, taking account differing
tunnelling rates to different orbitals.  Two distinct types of
coherence became evident from the results, a Hilbert space coherence
between states with same particle number, and a Fock-space coherence
between states with different particle numbers.  Under appropriate
conditions, the Fock-space coherence is decoupled from the evolution
of the rest of the system, suggesting promising consequences for
quantum information processing.

\ack 

This work is supported by the Natural Sciences and Engineering
Research Council of Canada, by the Canadian Foundation for Innovation,
and by Dalhousie University.  The authors acknowledge illuminating
discussions with Jean-Marc Samson and Catherine Stevenson.

\section*{References}

\bibliography{kyriakidis_birs_ldsn}
\bibliographystyle{iopart-num}

\end{document}